\documentclass[pra,
amssymb,amsmath,amsmath,showpacs,reprint,twocolumn]{revtex4-1}
\usepackage{color}
\usepackage{graphicx}
\usepackage{epstopdf}
\usepackage{natbib}
\usepackage{wasysym}
\usepackage{hyperref}
\usepackage{dcolumn}

\hypersetup{%
   pdfpagemode=None, 
   pdfstartpage=1,
   pdfstartview=FitH,
   pdfmenubar=true,
   pdftoolbar=true,
   colorlinks = true,
   linkcolor=blue,
   citecolor=blue,
   bookmarksopen=false
 }

\newcommand{\Fkt}[1]{\,\mathsf {#1}}

\def\openone{\leavevmode\hbox{\small1\kern-3.3pt\normalsize1}}

\ifx\Tr\renewcommand{\Tr}{\Fkt{Tr}} 
\else\newcommand{\Tr}{\Fkt{Tr}}
\fi

\usepackage{booktabs}
\usepackage{graphicx}
\usepackage{amsmath}
\usepackage{indentfirst}
\begin{document}
\title{Calculation of two-centre two-electron integrals over Slater-type orbitals revisited. \\I.
Coulomb and hybrid integrals}

\author{\sc Micha\l\ Lesiuk}
\email{e-mail: lesiuk@tiger.chem.uw.edu.pl}
\author{\sc Robert Moszynski}
\affiliation{\sl Faculty of Chemistry, University of Warsaw\\
Pasteura 1, 02-093 Warsaw, Poland}
\date{\today}

\begin{abstract}
In this paper, which constitutes the first part of the series, we consider calculation of two-centre Coulomb and hybrid 
integrals over Slater-type orbitals (STOs). General formulae for these integrals are derived with no restrictions 
on the values of the quantum numbers and nonlinear parameters. Direct integration over the coordinates of one of the 
electrons leaves us with the set of overlap-like integrals which are evaluated by using two distinct methods. The first 
one is based on the transformation to the ellipsoidal coordinates system and the second utilises a recursive scheme for 
consecutive increase of the angular momenta in the integrand. In both methods simple one-dimensional numerical 
integrations are used in order to avoid severe digital erosion connected with the straightforward use of the 
alternative analytical formulae. It is discussed that the numerical integration does not introduce a large 
computational overhead since the integrands are well-behaved functions, calculated recursively with decent speed. 
Special attention is paid to the numerical stability of the algorithms. Applicability of the resulting scheme over a 
large range of the nonlinear parameters is tested on examples of the most difficult integrals appearing in the actual 
calculations including at most 7$i$-type functions ($l$=$6$).
\end{abstract}

\maketitle

\section{Introduction}
\label{sec:intro}
Slater-type orbitals \cite{slater30,slater32}, or more general exponential-type orbitals, are the natural choice of
basis set for applications in quantum chemistry and molecular or atomic physics. Their common origin is the analytical
solution of the Schr\"{o}dinger equation for the hydrogen atom. It can be shown that Slater-type orbitals
behave correctly at the electron-nucleus coalescence points \emph{i.e.} they satisfy Kato's conditions
\cite{kato57}. Additionally, the Slater-type orbitals decay exponentially when an electron is far from the nucleus. This
is in line with the theoretical findings of the asymptotic form of the electron density \cite{agmon82}. It is obvious
that Gaussian orbitals \cite{boys50}, which have gained an enormous popularity in the last 50 years, are able to satisfy
neither of the above conditions. Virtually the only issue which prohibited the widespread use of the Slater-type
orbitals is calculation of the two-electron molecular integrals. 

The main purpose of the present series of papers is to provide a complete set of methods for the evaluation of the
two-electron two-centre integrals. The reliability of these methods needs to be sufficient to allow the use of 
Slater-type orbitals including high angular momentum functions for the diatomic systems. Our integral
program based on the presented algorithms serves as a vehicle for the upcoming new \emph{ab-initio} quantum chemistry
program package \textsc{Ko\l os}. This program combines basis set of Slater-type orbitals with state-of-the-art
quantum chemical \emph{ab-initio} methods and is aimed at spectroscopically accurate (few $\mbox{cm}^{-1}$) 
results for the diatomic systems.

When considering our approach to the present problem, one issue needs to be clarified. To reach the spectroscopic
accuracy it is not only necessary to use huge basis sets but also very
accurate quantum chemistry methods. Let us now observe that calculations of the two-electron integral file scale as
the fourth power of the size of the system ($N^4$) in the worst case scenario. This can be compared with the
scaling of the accurate coupled cluster methods, $N^6$ for CCSD, $N^8$ for CCSDT \emph{etc.}
\cite{bartlett89,raghavachari89,noga87,bartlett90}. As a result, one can
expect that calculations of the integral file should not be a bottleneck in high-level calculations of the correlation
energy. On the other hand, since we require the aforementioned accuracy in the molecular energy, we need the integrals
to be calculated with higher precision than typical. We believe that the requirement for accuracy of 12 decimal places
is reasonable.

The situation described above suggests that we should favour accuracy of the algorithms over their speed. In other
words, if we had two algorithms - the first one being fast but less accurate, and the second one being somehow slower
but significantly more accurate - we would pick up the second one. Of course, we still have limitations on the
computational time and we cannot use arbitrary precision arithmetic, for instance. This philosophy of choosing and
developing algorithms shall be perceptible throughout the whole series of papers.

This series of papers is organised as follows. In Paper I we deal with calculation of the Coulomb and hybrid 
integrals \emph{i.e.} $(aa|bb)$ and $(aa|ab)$, respectively, where $a$ and $b$ denote the nuclei at which orbitals are 
located. We use direct integration over the second electron in the same spirit as several previous investigators but 
we differ in methods of computation of almost all nontrivial basic 
quantities. Final forms of the working expressions are also completely reformulated. Moreover, we present the results of
demanding 
tests of the numerical performance. In Paper II we apply the Neumann expansion to calculation of the exchange 
integrals, $(ab|ab)$. We report new methods of calculation of the most difficult auxiliary quantities appearing in the 
theory. Additionally, we discuss how new algorithms can be sewed together to form a sufficiently general method. 
Finally, in Paper III we provide the first application of the presented theory - \emph{ab-initio} calculations for the 
beryllium dimer which is an interesting system from both spectroscopic and theoretical point of view. In these 
calculations we use STO basis sets ranging from double to sextuple zeta quality combined with high level 
\emph{ab-initio} methods in order to provide spectroscopically accurate results.

The literature dealing with evaluation of the molecular integrals over Slater-type orbitals is extensive
and a full bibliography would count hundreds of positions. Its detailed review is undoubtedly beyond the scope
of the present report. Therefore, our introduction is, by necessity, limited and subjective. Nonetheless, let us recall
several prominent and the most widely used general techniques for computation of the aforementioned integrals.

Single-centre expansions allow to expand STOs located at some point of space around a different centre. These methods
were pioneered by Barnett and Coulson as the widely known $\zeta$-function method
\cite{coulson37,barnett51,barnett63} and later independently by L\"{o}wdin \cite{lowdin56} ($\alpha$-function method).
In cases when the single-centre expansion terminates under the integral sign due to spherical symmetry of
the integrand, it typically results in closed-form, compact and plausible expressions. However, in many cases, such as
calculation of the exchange integrals, the single-centre expansions results in an infinite series which have a
pathologically slow
(\emph{i.e.} logarithmic) convergence rate \cite{flygare66}. The problem does not have a satisfactory solution
although several approaches \cite{barnett90} were adopted to overcome it. The second problem of the single-centre
expansions is the catastrophic
digital erosion during calculations of the auxiliary quantities \cite{jones92,kennedy99} which seems to be extremely
difficult to overcome. A promising work-around is use of the symbolic computational environments such as
\textsc{Mathematica} \cite{jones94,barnett00,barnett002} but at present the symbolic methods are typically orders of
magnitude slower than the numerical ones. Since the time the single-centre methods were first proposed, several new (or 
more general) expansion techniques has been developed. Examples are the works of Guseinov \cite{guseinov03}, Harris and 
Michels \cite{harris65} and Rico \emph{et al.} \cite{rico86,rico90} and references therein.

The second class of methods which gained a significant interest is the Gaussian expansion methods and the Gaussian
transform methods. The former is simply based on a least squares fit of a linear combination of Gaussian orbitals in
order to mimic the shape of STOs. This idea, proposed first by Boys and Shavitt, \cite{boys56} was the dominant method 
used in the early versions of the SMILES program \cite{smiles}. The Gauss transform methods are more involved and use 
some integral representations in order to transform STO into a more computationally convenient form. The initial 
proposition of Shavitt and Karplus \cite{shavitt62,shavitt65,kern65} was to use the Laplace transform of the 
exponential function but now a handful of different schemes is in use, along with suitable discretisation techniques 
\cite{rico04}.

The next prominent technique is the family of Fourier-transform methods which are usually used in conjunction with the
so-called $B$-functions. These methods were primarily developed by the group of Steinborn \cite{steinborn78, 
steinborn83a,steinborn83b,steinborn83c,steinborn86a,steinborn86b,steinborn88a,steinborn88b,steinborn90,steinborn91,
delhalle80} and applied to many difficult cases of the many-centre integrals. The fact that $B$-functions, being 
essentially a linear combination of STOs, possess an exceptionally simple Fourier transform can be used to evaluate the 
integrals in the momentum space and reduce many important integrals to the combination of some one-dimensional 
integrals. However, these integrals contain highly oscillatory integrands (including the Bessel functions) which make 
numerical integration extremely difficult with standard Gaussian quadrature techniques. Some approaches were adopted to 
accelerate the convergence towards the exact value with increasing number of quadrature nodes. The prominent method is 
the SD-transform, put forward by Sidi \cite{sidi81,sidi82}, and later applied by Safouhi \cite{safouhi04,safouhi06}. 
Despite that, it seems that there is no general method reliable enough to evaluate the integrals in question in a 
black-box fashion.

There is also a number of less extensively studied techniques for evaluation of the molecular integrals over STOs.
These include the Coulomb Sturmians introduced by Shull and L\"{o}wdin \cite{shull59} and used by some other authors 
\cite{smeyers66,guseinov02,avery04,red04}. The shift operator technique \cite{rico00a,rico00b,rico01} is a very elegant 
method which generates integrals with arbitrary STOs starting with the simplest integrals with 1s functions by 
application of the so-called shift operator. Gill \emph{et al.} \cite{gill08,gill09} introduced the Coulomb resolution 
techniques where the interaction potential is expanded in terms of the so-called potential functions resulting from the
the Poisson equation. This method has been recently pursued by Hoggan and coworkers \cite{hoggan09,hoggan10} and
included in their STOP program package \cite{stop}.

Remarkably, it has not been a well-known fact yet that all two-centre integrals over STOs were integrated analytically 
in a closed-form. In a recent work Pachucki \cite{pachucki09,pachucki12a} has shown that the so-called master 
integral with inverse powers of all interparticle distances can be obtained from the second order differential equation 
in the distance between the nuclei. The present authors also contributed to the development of this theory by extending 
it to the case of Slater geminals \cite{lesiuk12}. Pachucki used these expressions for calculations of the 
Born-Oppenheimer potential for the hydrogen molecule \cite{pachucki10,pachucki13} and helium hydride ion
\cite{pachucki12b}. However, an extreme level of complication of this theory along with drastic numerical 
instabilities occurring in the calculations have made its use limited to certain special forms of the basis set,
applicable only to two-electron systems. We believe that some 
ingenious reformulation of this theory is necessary to circumvent the aforementioned difficulties.

We postpone the discussion of the methods based on the Neumann expansion of the interaction potential in 
the ellipsoidal coordinates. In the second paper of the series it is used to evaluate the exchange integrals and a 
proper separate introduction is given therein.

Let us now concentrate on methods designed specifically for treatment of the Coulomb and hybrid integrals. For
the former ones there exists a plethora of independent methods which differ in both accuracy and speed. Probably the 
first attack on this problem was attempted by Barnett and Coulson \cite{coulson37} by using the 
single-centre expansion technique. Roothaan \cite{roothaan51} pioneered the direct integration method in the 
ellipsoidal coordinates which was later pursued by several authors 
\cite{ruedenberg56,roothaan56,ruedenberg66,silver68a,silver68b,ruedenberg69,eschrig79}. Later, it has become apparent 
that integration in the momentum space utilising the Fourier representation of STOs is very advantageous 
\cite{geller64,szondy65,silverstone67,silverstone70,harris69,steinborn83c,steinborn86a}. Gaussian transform techniques 
\cite{shavitt62,shavitt65,rico88}, refined translation/expansion methods \cite{sharma76,silverstone77,pendls91} and 
several special approaches \cite{rico88,tai92,jones93,hierse93,hierse94,magnasco89,magnasco90} were also 
successfully applied. For hybrid integrals the number of available methods is modest. Several prominent techniques,
such as the 
Fourier transform, cannot be applied straightforwardly. The biggest effort was aimed 
at the direct integration \cite{ruedenberg56,ruedenberg67,ruedenberg68,miller68} or its combinations with the 
translation techniques \cite{guseinov77,guseinov83,guseinov89}. Our unified approach to the Coulomb and hybrid 
integrals is based on the earlier experiences with the direct integration. By using the Laplace expansion of the 
interaction potential and 
analytic integration over the coordinates of the second electron the problem is reduced to the calculation of the 
standard 
overlap integrals and a set of \emph{overlap-like} integrals. To calculate these integrals two distinct approaches are 
used. The first one is integration in the ellipsoidal coordinates and the second method is based on recursive 
techniques. In both cases a simple, one-dimensional numerical integration is used to avoid drastic digital erosion. 
This indicates some connections with the method of Miller \cite{miller68}. Finally, we verify that when both methods 
are used together, in their respective regions of applicability, a loss of digits observed in the calculations by using 
some other methods can be avoided within a reasonable range of the nonlinear parameters.

Let us also note in passing that to perform actual calculations on the diatomic systems one also requires  
one-electron two-centre and two-electron one-centre (atomic) integrals. The former can be computed using 
various 
techniques among which the Fourier transform methods 
\cite{steinborn78,steinborn83a,steinborn83b,steinborn83c,steinborn86a,steinborn86b,steinborn88a,steinborn88b,
steinborn90,steinborn91,delhalle80}, recursive techniques for increasing the angular momenta in the integral 
\cite{guseinov05,guseinov07,rico88,rico89b,rico93,rico91,rico89c}, and finally direct integration using the ellipsoidal 
coordinate system \cite{mulliken49,corbato56,roothaan51,harris60} were intensively studied. The latter seems to be the 
method of choice for these integrals. Two-electron atomic integrals have been solved at least since the papers of 
Clementi and co-workers (see Refs. \cite{clementi74} and references therein). For the sake of completeness, a 
refined, simple, and numerically stable procedure for the computation of these integrals was included in the 
Supplementary Material \cite{supplement}.

\section{Preliminaries} 
\label{sec:pre}
Let us consider a diatomic system with the nuclei A and B centred at the positions $\textbf{R}_A=(0,0,-R/2)$ and
$\textbf{R}_B=(0,0,R/2)$, respectively, in the ordinary Cartesian coordinate system. Slater-type orbitals (STOs) have
the following general form:
\begin{align}
\label{sto1}
\chi_{nlm}(\textbf{r};\zeta)=S_n(\zeta) r^{n-1} e^{-\zeta r} Y_{lm}(\theta,\phi),
\end{align}
Therefore, any STO is uniquely described by the quartet of parameters $(n,l,m,\zeta)$. We assume throughout that $n$,
$l$
are restricted to the positive integers ($n>l$). The variables $r_a$, $\theta_a$,
$\phi$ denote the spherical coordinates located on the atom $A$ with analogous notation for the centre $B$. In Eq.
(\ref{sto1}), $S_n(\zeta)$ is the radial normalisation constant:
\begin{align}
S_n(\zeta)=\frac{(2\zeta)^{n+1/2}}{\sqrt{(2n)!}},
\end{align}
and $Y_{lm}(\theta,\phi)$ are the spherical harmonics defined according to the Condon-Shortley phase
convention \cite{messiah99}:
\begin{align}
Y_{lm}(\hat{\bf r}) = \Omega_{lm} P_l^{|m|}(\cos \theta)\frac{e^{\dot{\imath}m\phi}}{\sqrt{2\pi}},
\end{align}
where $P_l^m$ are the (unnormalised) associated Legendre polynomials \cite{stegun72} and $\Omega_{lm}$ is the angular
normalisation constant:
\begin{align}
\Omega_{lm} = \dot{\imath}^{m-|m|} \sqrt{\frac{2l+1}{2}\frac{(l-|m|)!}{(l+|m|)!}}
\end{align}
In actual calculations it is typical to use real versions of the spherical harmonics. However, the complex spherical
harmonics are more convenient in the derivations and thus we use them throughout the paper. Transfer to the real
spherical harmonics can be performed on the top of the presented algorithms by using standard relations.

Let us now introduce the prolate ellipsoidal coordinates $(\xi,\eta,\phi)$ by means of the following relations:
\begin{align}
\xi  = \frac{r_a+r_b}{R},\;\; \eta = \frac{r_a-r_b}{R},
\end{align}
so that $1\leq \xi \leq \infty$, $-1 \leq \eta \leq 1$ and $0 \leq \phi \leq 2\pi$. The spherical coordinates
are expressed through the ellipsoidal coordinates by means of the well-known expressions
\begin{align}
&r = \frac{R}{2}\left(\xi+\kappa\eta\right),\;\;
\cos \theta = \frac{1+\kappa\xi\eta}{\xi+\kappa\eta},
\end{align}
where the value of $\kappa$ is equal to $+1$ if STO is located on the centre A or $-1$ if it is located on the centre B.
The volume element becomes $d\textbf{r} = \left(\frac{R}{2}\right)^3(\xi^2-\eta^2)\,d\xi\,d\eta\,d\phi$. The simplest
way to express the product of two STOs (\emph{i.e.} the charge distribution) in the ellipsoidal coordinates is to
proceed
in two steps. First, we transfer the following scaled product of the Legendre polynomials by means of the expression
\begin{align}
\label{trans1}
\small
\begin{split}
&P_{l_a}^{|m_a|}(\cos \theta_a)P_{l_b}^{|m_b|}(\cos
\theta_b)r_a^{l_a}r_b^{l_b}=\\
&\left(\frac{R}{2}\right)^{l_a+l_b}
\left[(\xi^2-1)(1-\eta^2)\right]^{|M|/2}
\sum_{p=0}^{\Gamma}\sum_{q=0}^{\Gamma} \left( {\bf \large \Xi}_{l_al_b}^M
\right)_{pq} \xi^p \eta^q
\end{split}
\end{align}
where $M=m_a-m_b$, $\Gamma=l_a+l_b$ and ${\bf \large \Xi}_{l_al_b}^M$ are square matrices of dimension
$\Gamma$. The values of the latter depend on the locations of the orbitals and their quantum numbers. Explicit forms of
these matrices can easily be deduced from the general expressions available in the literature
\cite{maslen90,rico89,yasui82,guseinov70,guseinov97}. We tabulated the values of ${\bf \large \Xi}_{l_al_b}^M$ up
to the maximum value of
$l_a+l_b$ equal to 24. These tables, along with \textsc{Mathematica} code \cite{math7} used for their generation, can be
obtained from the authors on demand.

The remainder can be transferred to the ellipsoidal coordinates by using the following formula:
\begin{align}
\label{trans2}
r_a^{n_a}r_b^{n_b}=\sum_{k=0}^{k_{max}} B_k^{n_an_b}\xi^k \eta^{k_{max}-k},
\end{align}
with $k_{max} = n_a+n_b$. The above expression has been extensively used by many authors
\cite{guseinov70,rosen31,ozdogan04} who presented explicit
expressions for the coefficients $B_k^{n_an_b}$ (the so-called generalised binomial coefficients). We found it simpler
to tabulate these coefficients as series of
one-dimensional look-up tables. 

Making use of the transfer relations (\ref{trans1}) and (\ref{trans2}) one can write down the explicit expression for
the STOs charge distribution in terms of the ellipsoidal coordinates. The result reads (for convenience, we
additionally included the Jacobian):
\begin{align}
\label{trans12}
\begin{split}
&\left(\frac{R}{2}\right)^3(\xi^2-\eta^2)\chi_{n_al_am_a}^*(\textbf{r}_a;\zeta_a)\chi_{n_bl_bm_b}(\textbf{r}
_b;\zeta_b)=\\
&\frac{K_{ab}}{2\pi}e^{-\alpha\xi-\beta\eta} \left[(\xi^2-1)(1-\eta^2)\right]^{|M|/2}e^{\dot{\imath}M\phi}\times\\
&\sum_{k=0}^{k_{max}} B_k^{n_a-l_a,n_b-l_b}\sum_{p=0}^{\Gamma}\sum_{q=0}^{\Gamma} \left( {\bf \large \Xi}_{l_al_b}^M
\right)_{pq} \xi^{p+k}\; \eta^{q+k_{max}-k},
\end{split}
\end{align}
with $M=m_a-m_b$, $k_{max}=n_a-l_a+n_b-l_b$ and $\Gamma=l_a+l_b$. Additionally, in Eq. (\ref{trans12}) we introduced
several new quantities: $\alpha=\frac{R}{2}(\zeta_a+\zeta_b)$,
$\beta=\frac{R}{2}(\kappa_a\zeta_a+\kappa_b\zeta_b)$,
$K_{ab}=S_{n_a}(\zeta_a) S_{n_b}(\zeta_b)\, \Omega_{l_am_a} \Omega_{l_bm_b} \left(\frac{R}{2}\right)^{n_a+n_b+1}$. The
above formulation is quite explicit and rather transparent at the same time. Apart from that, it remains valid for
``singular'' orbitals such as $0s$ which is advantageous from the point of view of some developments.

Before passing further let us introduce three useful auxiliary functions:
\begin{align}
\label{Ap}
A_p(\alpha) = \int_1^\infty d\xi\; \xi^p e^{-\alpha \xi},
\end{align}
\begin{align}
\label{Bq}
B_q(\beta ) = \int_{-1}^{+1} d\eta\; \eta^q e^{-\beta \eta},
\end{align}
\begin{align}
\label{ap}
a_p(\alpha) = \int_0^1 d\xi\; \xi^p e^{-\alpha \xi}.
\end{align}
The first two of the above functions are the so-called Mulliken integrals \cite{mulliken49}. Accurate and stable
calculation of these integrals was considered by many authors, the works of Corbat\'{o} \cite{corbato56} and a 
recent paper of Harris \cite{harris04} need to be mentioned in this respect. The third integral, Eq. (\ref{ap}), can be
considered complementary to the first integral, Eq. (\ref{Ap}), and has strong connections with the lower incomplete
gamma functions. Integrals (\ref{ap}) have to be computed by using the Miller algorithm \cite{gautschi67}, as discussed
by Harris \cite{harris02}.

\section{Coulomb and hybrid integrals}
\label{sec:coulomb}
In this section we attack the main objectives of this paper - calculation of the coulomb ($I_C$) and hybrid ($I_H$)
integrals. With the notation developed in the previous section they take the following form:
\begin{align}
\begin{split}
I_C = \int d\textbf{r}_1 \int d\textbf{r}_2\,
\chi_{n_1l_1m_1}^*(\textbf{r}_{1a};\zeta_1) \chi_{n_2l_2m_2}(\textbf{r}_{1a};\zeta_2)\\
\times\frac{1}{r_{12}}\,\chi_{n_3l_3m_3}^*(\textbf{r}_{2b};\zeta_3) \chi_{n_4l_4m_4}(\textbf{r}_{2b};\zeta_4),\\
\end{split}
\end{align}
\begin{align}
\begin{split}
I_H = \int d\textbf{r}_1 \int d\textbf{r}_2\,
\chi_{n_1l_1m_1}^*(\textbf{r}_{1a};\zeta_1) \chi_{n_2l_2m_2}(\textbf{r}_{1b};\zeta_2)\\
\times\frac{1}{r_{12}}\,\chi_{n_3l_3m_3}^*(\textbf{r}_{2b};\zeta_3) \chi_{n_4l_4m_4}(\textbf{r}_{2b};\zeta_4).
\end{split}
\end{align}
Let us note that in the above expressions we have adapted a particular, fixed location of the STOs. This
convention is very useful from the point of view of the upcoming derivation. Other possible options for the
orbitals location within the class of the Coulomb and hybrid integrals can be obtained by using the usual eightfold
permutational symmetry of the integrals.

\subsection{Initial reduction to the overlap-like integrals}
\label{subsec:chinit}
Before proceeding with the integration of $I_C$ and $I_H$ let us simplify the formulae by using the Clebsh-Gordan
expansion of the products of the spherical harmonics. In the case of the Coulomb integrals one expands pairs of the
spherical harmonics on both centres; in case of the hybrid integrals, only the pair dependent on the coordinates of the
second electron can be expanded. Once the Clebsh-Gordan expansion is used and the resulting integrals are written
explicitly, it becomes obvious that the problem reduces now to the calculation of the following families of the
integrals
\begin{align}
\begin{split}
\widetilde{I}_C = &\int d\textbf{r}_1 \int d\textbf{r}_2\,
r_{1a}^{n_{12}-2}\, Y_{L_1M}^*(\cos \theta_{1a},\phi) 
\,\frac{1}{r_{12}}\\
&\times r_{2b}^{n_{34}-2}\, Y_{L_2M}(\cos \theta_{2b},\phi)\, e^{-\zeta_{12}r_{1a}-\zeta_{34}r_{2b}},
\end{split}
\end{align}
\begin{align}
\begin{split}
\widetilde{I}_H = &\int d\textbf{r}_1 \int d\textbf{r}_2\, 
r_{1a}^{n_1-1}\, Y_{l_1m_1}^*(\cos \theta_{1a},\phi)e^{-\zeta_1r_{1a}}\\
&\times
r_{1b}^{n_2-1}\, Y_{l_2m_2}  (\cos \theta_{1b},\phi)e^{-\zeta_2r_{1b}}
\,\frac{1}{r_{12}}\\
&\times r_{2b}^{n_{34}-2}\, Y_{L_2M}(\cos \theta_{2b},\phi)\, e^{-\zeta_{34}r_{2b}},
\end{split}
\end{align}
where $n_{12}=n_1+n_2$, $\zeta_{12}=\zeta_1+\zeta_2$ \emph{etc}. and $r_{ij}$ denote the interparticle distances. It is
evident that any Coulomb integral ($I_C$) can be written as a linear combination of the pertinent integrals
$\widetilde{I}_C$ and the correspondence between $I_H$ and $\widetilde{I}_H$ is analogous. For convenience, we have
also skipped the normalisation constants $S_n$ since their multiplicative presence is obvious and does not change
throughout the derivation. When considering the coefficients that relate $I_{C/H}$ and $\widetilde{I}_{C/H}$ there is an
additional effort connected with calculation of the Wigner $3J$ symbols (or equivalently the Clebsh-Gordan 
coefficients).
Computation of these quantities is not a trivial problem and has been considered many times in the literature,
see Refs. \cite{venkatesh78,tuzun98,schulten76,fang92,lai90,rasch03} as representative examples.

The first step of the integration proceeds in the usual manner - one integrates over the coordinates of the second
electron. This is a quite natural approach since both orbitals of the second electron lie on the same centre (B, in our
convention). The formula for the necessary integral exists in the literature and appears independently in many works.
The simplest way to arrive at the final expression is to use the Laplace expansion of $1/r_{12}$ in spherical
coordinates relative to the centre B. Independently of the derivation route, one arrives at
\begin{widetext}
\begin{align}
\label{int2}
\begin{split}
\int d\textbf{r}_2\,\frac{1}{r_{12}} r_{2b}^{n_{34}-2}\, Y_{L_2M}(\cos \theta_{2b},\phi)\, e^{-\zeta_{34}r_{2b}}&=
\frac{4\pi}{2L_2+1}\frac{Y_{L_2M}(\cos \theta_{1b},\phi)}{\zeta_{34}^{n_{34}}}
\bigg[ (\zeta_{34}r_{1b})^{n_{34}}\,a_{n_{34}+L_2}(\zeta_{34}r_{1b})\\
&+(n_{34}-L_2-1)!\;e^{-\zeta_{34}r_{1b}}\sum_{j=L_2}^{n_{34}-1}\frac{(\zeta_{34}\,r_{1b})^j}{(j-L_2)!}\bigg],
\end{split}
\end{align}
\end{widetext}
where $a_n$ is given by Eq. (\ref{ap}). To bring the above expression into a more familiar and simplified form we could
use the following obvious relationships
\begin{align}
\label{illegal1}
a_n(\alpha) &= \frac{n!}{\alpha^{n+1}} - A_n(\alpha),
\end{align}
\begin{align}
\label{illegal2}
A_n(\alpha) &= \frac{e^{-\alpha}\, n!}{\alpha^{n+1}} \sum_{k=0}^n \frac{\alpha^k}{k!}.
\end{align}
By doing so, one expresses the integral (\ref{int2}) explicitly through the elementary functions only. It seems to be
advantageous but there are two main problems connected with use of Eqs. (\ref{illegal1}) and (\ref{illegal2}). Firstly,
these expressions introduce spurious singularities (high inverse powers of $r_{1b}$) and generate integrals which have
to be treated with special methods. Secondly, and more importantly, Eq. (\ref{illegal1}) by itself is numerically
badly conditioned and these problems propagate to the final expressions for the Coulomb and hybrid integrals. 
Precisely
speaking, unless the relationship $n\gg \alpha$ holds, Eq. (\ref{illegal1}) consists of subtraction of two large
numbers to a relatively small result. Therefore, a huge digital erosion occurs, especially when large values of the
quantum numbers are necessary.

This leads to the conclusion that in order to preserve a good numerical stability of the method, we have to abandon the
use of Eqs. (\ref{illegal1}) and (\ref{illegal2}) and exploit Eq. (\ref{int2}) as it stands. By inserting Eq.
(\ref{int2}) into the initial expressions for $\widetilde{I}_{C}$ one obtains the formula
\begin{widetext}
\begin{align}
\label{ic}
\small
\begin{split}
\widetilde{I}_{C} &= \frac{4\pi}{2L_2+1} \frac{1}{\zeta_{34}^{n_{34}}}
\bigg[\zeta_{34}^{n_{34}} \int d\textbf{r}_1 \;r_{1a}^{n_{12}-2} \;Y_{L_1M}(\cos
\theta_{1a},\phi)\;e^{-\zeta_{12}r_{1a}}\;
r_{1b}^{n_{34}}\;Y_{L_2M}(\cos \theta_{1b},\phi)\;a_{n_{34}+L_2}(\zeta_{34}r_{1b})\\
&+ (n_{34}-L_2-1)! \sum_{j=L_2}^{n_{34}-1} 
\frac{\zeta_{34}^j}{(j-L_2)!}
\int d\textbf{r}_1 \;r_{1a}^{n_{12}-2} \;Y_{L_1M}(\cos
\theta_{1a},\phi)\;e^{-\zeta_{12}r_{1a}}\;
r_{1b}^j\;Y_{L_2M}(\cos \theta_{1b},\phi)\;e^{-\zeta_{34}r_{1b}} \bigg].
\end{split}
\end{align}
\end{widetext}
For the hybrid integrals, the manipulations are slightly more involved. After inserting Eq. (\ref{int2}) into the 
formula
for $\widetilde{I}_{H}$ one is left with three spherical harmonics under the integral sign. Two of these spherical
harmonics are centred at the nucleus B and therefore can be expanded in the Clebsh-Gordan series. The result of this
manipulations is as follows (the usual notation for the Wigner $3J$ symbols is used):
\begin{widetext}
\begin{align}
\label{ih}
\small
\begin{split}
\widetilde{I}_{H} &= \frac{(-1)^{m_2}}{\zeta_{34}^{n_{34}}}\sqrt{(2l_2+1)(2L_2+1)} 
\sum_{L_1=|l_2-L_2}^{l_2+L_2} \sqrt{\frac{4\pi}{2L_1+1}}
\left(\begin{array}{cccc} l_2 & L_2 & L_1\\ -m_2 & -M & m_1 \end{array}\right)
\left(\begin{array}{cccc} l_2 & L_2 & L_1\\ 0 & 0 & 0 \end{array}\right)\\
&\times \bigg[\zeta_{34}^{n_{34}} \int d\textbf{r}_1 \;r_{1a}^{n_1-1} \;Y_{l_1m_1}(\cos
\theta_{1a},\phi)\;e^{-\zeta_{1}r_{1a}}\;
r_{1b}^{n_{2}+n_{34}-1}\;Y_{L_1m_1}(\cos \theta_{1b},\phi)\;e^{-\zeta_2r_{1b}}\;a_{n_{34}+L_2}(\zeta_{34}r_{1b})\\
&+ (n_{34}-L_2-1)! \sum_{j=L_2}^{n_{34}-1} 
\frac{\zeta_{34}^j}{(j-L_2)!}
\int d\textbf{r}_1 \;r_{1a}^{n_{1}-1} \;Y_{l_1m_1}(\cos
\theta_{1a},\phi)\;e^{-\zeta_{1}r_{1a}}\;
r_{1b}^{n_2+j-1}\;Y_{L_1m_1}(\cos \theta_{1b},\phi)\;e^{-(\zeta_2+\zeta_{34})r_{1b}} \bigg].
\end{split}
\end{align}
\end{widetext}
Let us now investigate the above formulae in a greater detail. It is obvious that Eqs. (\ref{ic}) and (\ref{ih}) 
include two basic types of integrals which take the following general forms
\begin{widetext}
\begin{align}
\label{ovrlp1}
S_{n_1l_1m}^{n_2l_2m}(\zeta_1,\zeta_2) = 
\int d\textbf{r}_1 \;r_{1a}^{n_1-1} \;Y_{l_1m}(\cos
\theta_{1a},\phi)\;e^{-\zeta_1r_{1a}}\;
r_{1b}^{n_2-1}\;Y_{l_2m}(\cos \theta_{1b},\phi)\;e^{-\zeta_2r_{1b}},
\end{align}
\begin{align}
\label{ovrlp2}
\widetilde{S}_{n_1l_1m}^{n_2l_2m}({n_3};\zeta_1,\zeta_2,\zeta_3) = 
\int d\textbf{r}_1 \;r_{1a}^{n_1-1} \;Y_{l_1m}(\cos
\theta_{1a},\phi)\;e^{-\zeta_1r_{1a}}\;
r_{1b}^{n_2-1}\;Y_{l_2m}(\cos \theta_{1b},\phi)\;e^{-\zeta_2r_{1b}}\;a_{n_3}(\zeta_3r_{1b}),
\end{align}
\end{widetext}
The first integral is simply an overlap integral between two-centre STO charge distributions and for the second one let 
us introduce the name \emph{overlap-like integral}. The latter differs from the former only 
by the presence of $a_n$ function under the integral sign. Further, we concentrate solely on the 
overlap-like integrals and present two separate approaches. We shall verify that these two methods combined provide 
sufficient accuracy and reasonable speed to allow calculation of the desired Coulomb and hybrid integrals. We see no 
need to consider overlap integrals (\ref{ovrlp1}) separately. As one can see shortly, they can be computed by using 
exactly the same algorithms as integrals (\ref{ovrlp2}). The only differences lie in the fact that for the overlap-like 
integrals we use numerical integration to compute some of the basic quantities and for the overlap 
integrals, Eq. (\ref{ovrlp1}), this numerical integration can simply be skipped due to absence of the $a_n$ factor.

\subsection{Calculation of the overlap-like integrals by the ellipsoidal coordinates method}
\label{subsec:chinit}
For the calculation of the overlap-like integrals the use of ellipsoidal coordinates seems to be a natural approach 
because the
standard one-electron integrals separate into a product of simple one-dimensional integrals. It is obvious, however,
that due to the presence of the factor $a_{n}$ in Eq. (\ref{ovrlp2}) this separation can no longer be performed
straightforwardly. Not discouraged by this fact, we proceed in a
conventional manner and utilise Eq. (\ref{trans12}) to express the integrand in Eq. (\ref{ovrlp2}) in elliptic
coordinates. Noting that the axial symmetry of the integrand requires $M=0$ in the transfer formula (\ref{trans12}) we
arrive at the expression
\begin{widetext}
\begin{align}
\label{transs}
\widetilde{S}_{n_1l_1m}^{n_2l_2m}({n_3};\zeta_1,\zeta_2,\zeta_3) = 
K_{12}
\sum_{k=0}^{k_{max}} B_k^{n_1-l_1,n_2-l_2}\sum_{p=0}^{\Gamma}\sum_{q=0}^{\Gamma} \left( {\bf \large \Xi}_{l_1l_2}^0
\right)_{pq} \int_{+1}^\infty d\xi \int_{-1}^{+1} d\eta\; \xi^{p+k}\; \eta^{q+k_{max}-k}
e^{-\alpha\xi-\beta\eta}\;a_{n_3}\big[\gamma\big(\xi+\eta\big)\big],
\end{align}
\end{widetext}
after an elementary integration over the angle $\phi$. In the above expression
$K_{12}=(R/2)^{n_1+n_2+3}\,\Omega_{l1m1}\,\Omega_{l2m2}$, $k_{max}=n_1-l_1+n_2-l_2$, $\alpha$ and $\beta$ are defined
analogously as in Eq. (\ref{trans12}) and $\gamma=R\cdot\zeta_3/2$. Let us now consider the inner integrals in the
above expression and define the auxiliary integrals class:
\begin{align}
\label{jl1}
\begin{split}
J_\lambda(p,q;\alpha,\beta,\gamma)&= \int_{+1}^\infty d\xi \int_{-1}^{+1} d\eta\; \xi^p\; \eta^q \\
&\times e^{-\alpha\xi-\beta\eta}\;a_{\lambda}\big[\gamma\big(\xi+\eta\big)\big].
\end{split}
\end{align}
The above integrals do not separate to a product of one-dimensional integrals and are also very resistant to the
numerical integration. However, let us insert the integral representation (\ref{ap}) and change the order of
integration so that integrations over $\xi$ and $\eta$ are performed first. One easily recognises that the inner
integrals are the Mulliken integrals defined in Eqs. (\ref{Ap}), (\ref{Bq}) and the integrals (\ref{jl1}) can be
written as
\begin{align}
\label{jl2}
J_\lambda(p,q;\alpha,\beta,\gamma)= 
\int_0^1 dt\; t^\lambda A_p(\alpha+\gamma t)\;B_q(\beta-\gamma t).
\end{align}
Note that, apart from reducing the dimensionality of the integral, we have obtained a form which is very
convenient for the numerical integration. The Mulliken integrals are smooth, continuous functions of the
real variable with no singularities on the integration line or unwanted oscillatory behaviour. Therefore, there is no
need to use numerical quadratures with overwhelmingly large number of points. Additionally, the
Mulliken integrals can be calculated extremely efficiently in a recursive fashion for arbitrary values of the
parameters.

Despite the obvious advantages of the numerical integration of Eq. (\ref{jl2}) this approach still has to be justified
to some extent. One may ask what is the point of using numerical integration since integrals (\ref{jl2}) can be worked
out analytically. One can do that, for instance, by inserting in Eq. (\ref{jl2}) the explicit expressions for the
Mulliken integrals, which are available in the literature \cite{mulliken49}. Next, the integral over $t$ can be
expressed as a hypergeometric function of two integer parameters and with help of the so-called contiguous relations one
can reduce the initial integrals to combinations of the well-known basic functions. This approach seems to be
particularly attractive for the Coulomb integrals (when $\zeta_2=0$) since, as pointed out by Tai \cite{tai92}, the
final explicit expressions contain only elementary functions of the real variables. Therefore, the numerical approach to
the integrals (\ref{jl2}) seems to be an unwise decision at first glance.

However, the actual situation is more complicated. Taking Eq. (\ref{jl2}) as a starting point, we note that 
the explicit expressions for $B_q$ functions are badly conditioned due to cancellation of two large terms to a 
relatively small result. That is why computation of $B_q$ from the analytic expressions is unstable and alternate
methods need to be utilised \cite{corbato56,harris04}. This instability propagates further to the integrals (\ref{jl2}) 
and becomes more pronounced as the value of $q$ increases. Nonetheless, with help of the symbolic algebra package, such 
as \textsc{Mathematica}, one can derive explicit
expressions for $J_\lambda$ in order to verify their usefulness. We found that for $\beta\approx\gamma$ the loss of
digits is enormous, even when the values of $q$ are not large. Therefore, a prohibitively high arithmetic precision is 
required to obtain any useful information about the
values of $J_\lambda$. Taking into consideration the philosophy presented in the introduction (favouring accuracy over
speed within reasonable limits), the above observation seems to state a deadly argument against the analytic
approach. In other words, the numerical integration can be understood as a simple way to avoid a severe digital erosion.

\begin{table*}[ht]
\caption{Calculation of $\widetilde{S}_{13,12,m}^{15,12,m}({26};\zeta_1,0,\zeta_3)$ using the method based on
ellipsoidal
coordinates. The values of $\zeta_1$, $\zeta_3$ increase along the columns or rows, respectively. The values presented
are in the form d$-$q which denotes (rounded) decimal logarithms of the relative error obtained in double and
quadruple arithmetic precision, respectively. Therefore, this values roughly represent the number of correct significant
digits obtained using the present algorithm. Values obtained in quadruple precision were demoted to double precision
before the comparison since these are the values used in the actual calculations. The worst result obtained within the
possible range of $m$ was chosen in all cases. Reference values were obtained from calculations in extended arithmetic
precision of 128 significant digits. }
\begin{ruledtabular}
\begin{tabular}{c|ccccccccccccc}
\label{table1}
$\zeta_1$/$\zeta_3$ & 0.1250 & 0.2500 & 0.5000 & 1.0000 & 2.0000 & 4.0000 & 8.0000 & 16.000 & 32.000 & 64.000 & 128.00
& 
256.00\\
\hline\\[-2.1ex]
0.1250 & 9$-$16  & 9$-$16  & 8$-$16 & 9$-$16 & 8$-$16 & 7$-$16 & 8$-$16 & 5$-$16 & 1$-$16 & 0$-$13 & 0$-$5  & 0$-$0
\\[0.6ex]
0.2500 & 8$-$16  & 8$-$16  & 8$-$16 & 8$-$16 & 8$-$16 & 9$-$16 & 7$-$16 & 5$-$16 & 1$-$16 & 0$-$13 & 0$-$5  & 0$-$0
\\[0.6ex]
0.5000 & 9$-$16  & 8$-$16  & 8$-$16 & 8$-$16 & 8$-$16 & 7$-$16 & 6$-$16 & 5$-$16 & 1$-$16 & 0$-$12 & 0$-$4  & 0$-$0
\\[0.6ex]
1.0000 & 9$-$16  & 8$-$16  & 9$-$16 & 8$-$16 & 8$-$16 & 7$-$16 & 7$-$16 & 4$-$16 & 0$-$16 & 0$-$12 & 0$-$5  & 0$-$0
\\[0.6ex]
2.0000 & 8$-$16  & 8$-$16  & 8$-$16 & 9$-$16 & 8$-$16 & 7$-$16 & 8$-$16 & 5$-$16 & 0$-$16 & 0$-$11 & 0$-$4  & 0$-$0
\\[0.6ex]
4.0000 & 8$-$16  & 8$-$16  & 8$-$16 & 9$-$16 & 7$-$16 & 8$-$16 & 7$-$16 & 5$-$16 & 0$-$16 & 0$-$12 & 0$-$5  & 0$-$0
\\[0.6ex]
8.0000 & 8$-$16  & 8$-$16  & 8$-$16 & 6$-$16 & 7$-$16 & 8$-$16 & 7$-$16 & 5$-$16 & 1$-$16 & 0$-$14 & 0$-$7  & 0$-$1
\\[0.6ex]
16.000 & 7$-$16  & 7$-$16  & 7$-$16 & 5$-$16 & 5$-$16 & 5$-$16 & 6$-$16 & 7$-$16 & 4$-$16 & 0$-$16 & 0$-$9  & 0$-$3
\\[0.6ex]
32.000 & 4$-$16  & 4$-$16  & 4$-$16 & 3$-$16 & 2$-$16 & 2$-$16 & 1$-$16 & 4$-$16 & 6$-$16 & 0$-$16 & 0$-$11 & 0$-$4
\\[0.6ex]
64.000 & 0$-$16  & 0$-$16  & 0$-$15 & 0$-$14 & 0$-$15 & 0$-$14 & 0$-$14 & 0$-$16 & 2$-$16 & 1$-$16 & 0$-$12 & 0$-$4
\\[0.6ex]
128.00 & 0$-$10  & 0$-$10  & 0$-$8  & 0$-$8  & 0$-$7  & 0$-$7  & 0$-$8  & 0$-$11 & 0$-$14 & 1$-$15 & 0$-$12 & 0$-$4
\\[0.6ex]
256.00 & 0$-$4   & 0$-$4   & 0$-$2  & 0$-$1  & 0$-$1  & 0$-$0  & 0$-$1  & 0$-$4  & 0$-$7  & 0$-$14 & 0$-$12 & 0$-$4
\\[0.6ex]
\end{tabular}
\end{ruledtabular}
\end{table*}

For the benchmarking purposes, we show results of the calculation of two integrals,
$\widetilde{S}_{1312m}^{1512m}({26};\zeta_1,0,\zeta_3)$ and 
$\widetilde{S}_{7,6,m}^{21,18,m}({26};\zeta_1,\zeta_1,\zeta_3)$, within
the reasonable range of values of the nonlinear parameters $\zeta_1$, $\zeta_3$. We are free to set $R=1$ since
an increase of $R$ results only in scaling of the nonlinear parameters by $R$ (up to a trivial multiplicative 
constant). All necessary $J_\lambda$ integrals were calculated numerically using 100 or 200 grid points of the 
Tanh-Sinh quadrature \cite{tanhsinh1,tanhsinh2} for double and quadruple arithmetic precision, respectively. Under 
these 
conditions, $J_\lambda$ integrals are typically calculated with full precision allowed by the arithmetic. 

The integrals, $\widetilde{S}_{1312m}^{1512m}({26};\zeta_1,0,\zeta_3)$ and 
$\widetilde{S}_{7,6,m}^{21,18,m}({26};\zeta_1,\zeta_1,\zeta_3)$, are the most difficult quantities (in terms of the 
angular momentum) encountered in the calculation of the Coulomb and hybrid integrals, respectively, including at 
most 7$i$ functions. We set $\zeta_2 = \zeta_1$ in the second integral for illustrative purposes - the overall picture 
changes very slightly when the value of $\zeta_2$ is distorted. The results are presented in Table 1 for the first
integral and in Table 2 for the second integral. One observes a progressive loss of digits when one of the nonlinear 
parameters is large and the second is small. This digital erosion is due to the cancellation of large numbers during 
summations in Eq. (\ref{transs}) and it cannot be avoided in the ellipsoidal coordinates method. The use of 
quadruple precision improves the situation a lot but it is not sufficient to cope with the most difficult cases. Of
course, for 
lower angular momentum functions the changes are less sharp but the overall trend remains the same.
Concluding, our observations signal that the ellipsoidal coordinates method alone is not sufficient to calculate the 
desired integrals with the prescribed accuracy and need to be supplemented by a different algorithm.

In the present series of papers we do not go into technical details of the implementation \emph{etc.}, but
let us give a short remark on the timings in the present algorithm. The numerical integration of the integrals
$J_\lambda$ typically consumes about a half of the total time necessary to calculate a given shell of integrals. Only
for the smallest values of the quantum numbers this ratio is higher, but these integrals are very cheap anyway. The
remaining time is spent on the lengthy summations in Eq. (\ref{transs}), formation of $\widetilde{I}_{C/H}$, Eqs.
(\ref{ic}) and (\ref{ih}), and summation of the initial Clebsh-Gordan expansion to finally arrive at the value of
$I_{C/H}$. Therefore, the numerical integration is not connected with a drastic overhead as might have been initially
expected. A faster scheme for the calculation of $J_\lambda$ shall not result in a significant overall 
speed-up. Typically, the Coulomb and hybrid integrals are obtained in 1-100 $\mu$s per integral, depending on the values
of quantum numbers, with hybrid integrals being slightly more expensive.

\begin{table*}[ht]
\caption{Calculation of $\widetilde{S}_{7,6,m}^{21,18,m}({26};\zeta_1,\zeta_1,\zeta_3)$ using the method based on
the ellipsoidal
coordinates. The values
of $\zeta_1$, $\zeta_3$ increase along the columns or rows, respectively. The values presented
are in the form d$-$q which denotes (rounded) decimal logarithms of the relative error obtained in double and
quadruple arithmetic precision, respectively. Therefore, this values roughly represent the number of correct significant
digits obtained using the present algorithm. Values obtained in quadruple precision were demoted to double precision
before the comparison since these are the values used in the actual calculations. The worst result obtained within the
possible range of $m$ was chosen in all cases. Reference values were obtained from calculations in extended arithmetic
precision of 128 significant digits. }
\begin{ruledtabular}
\begin{tabular}{c|ccccccccccccc}
\label{table2}
$\zeta_1$/$\zeta_3$ & 0.1250 & 0.2500 & 0.5000 & 1.0000 & 2.0000 & 4.0000 & 8.0000 & 16.000 & 32.000 & 64.000 & 128.00 &
256.00\\
\hline\\[-2.1ex]
0.1250 & 0$-$9  & 0$-$10 & 0$-$12 & 0$-$16 & 1$-$16 & 6$-$16 & 7$-$16 & 7$-$16 & 6$-$16 & 6$-$16 & 4$-$16 & 4$-$16
\\[0.6ex]
0.2500 & 0$-$10  & 0$-$10 & 0$-$13 & 0$-$16 & 2$-$16 & 5$-$16 & 8$-$16 & 7$-$16 & 6$-$16 & 7$-$16 & 4$-$16 & 4$-$16
\\[0.6ex]
0.5000 & 0$-$10  & 0$-$10 & 0$-$13 & 0$-$16 & 3$-$16 & 6$-$16 & 7$-$16 & 7$-$16 & 6$-$16 & 7$-$16 & 4$-$16 & 4$-$16
\\[0.6ex]
1.0000 & 0$-$11 & 0$-$11 & 0$-$14 & 0$-$16 & 2$-$16 & 5$-$16 & 7$-$16 & 7$-$16 & 7$-$16 & 7$-$16 & 5$-$16 & 5$-$16
\\[0.6ex]
2.0000 & 0$-$12 & 0$-$13 & 0$-$15 & 0$-$16 & 2$-$16 & 5$-$16 & 7$-$16 & 6$-$16 & 6$-$16 & 6$-$16 & 5$-$16 & 5$-$16
\\[0.6ex]
4.0000 & 0$-$14 & 0$-$15 & 0$-$16 & 0$-$16 & 2$-$16 & 4$-$16 & 6$-$16 & 5$-$16 & 6$-$16 & 6$-$16 & 5$-$16 & 5$-$16
\\[0.6ex]
8.0000 & 0$-$16 & 0$-$16 & 0$-$16 & 0$-$16 & 2$-$16 & 3$-$16 & 4$-$16 & 4$-$16 & 4$-$16 & 4$-$16 & 7$-$16 & 5$-$16
\\[0.6ex]
16.000 & 0$-$16 & 0$-$16 & 0$-$16 & 0$-$16 & 1$-$16 & 1$-$16 & 1$-$16 & 2$-$16 & 2$-$16 & 2$-$16 & 4$-$16 & 3$-$16
\\[0.6ex]
32.000 & 0$-$13 & 0$-$14 & 0$-$13 & 0$-$14 & 0$-$15 & 0$-$15 & 0$-$15 & 0$-$16 & 0$-$16 & 0$-$16 & 0$-$15 & 0$-$14
\\[0.6ex]
64.000 & 0$-$6  & 0$-$6  & 0$-$6  & 0$-$7  & 0$-$7  & 0$-$8  & 0$-$9  & 0$-$9  & 0$-$9  & 0$-$6  & 0$-$5  & 0$-$4
\\[0.6ex]
128.00 & 0$-$0  & 0$-$0  & 0$-$0  & 0$-$0  & 0$-$0  & 0$-$0  & 0$-$1  & 0$-$2  & 0$-$2  & 0$-$4  & 0$-$3  & 0$-$2
\\[0.6ex]
\end{tabular}
\end{ruledtabular}
\end{table*}

\subsection{Calculation of the overlap-like integrals by the recursive method}

For the calculation of the overlap-like integrals by using the recursive method it is more convenient to introduce
different basic integrals, so that the final expressions take a simpler form. Let us note that Eq. (\ref{ovrlp2}) can
be rewritten as
\begin{align}
\widetilde{S}_{n_1l_1m}^{n_2l_2m}({n_3};\zeta_1,\zeta_2,\zeta_3) = 
\frac{1}{R}\,\Omega_{l_1m}\Omega_{l_2m} \langle \varphi_{n_1}^{l_1m}|\varphi_{n_2}^{l_2m}\rangle,
\end{align}
where
\begin{align}
\begin{split}
\label{pertinent1}
\langle \varphi_{n_1}^{l_1m}|\varphi_{n_2}^{l_2m}\rangle &=
 \int_0^\infty dr_a \int_{|r_a-R|}^{r_a+R} dr_b \;r_a^{n_1} r_b^{n_2}e^{-\zeta_1r_a-\zeta_2r_b}\\
&\times P_{l_1}^m(\cos \theta_a)\;P_{l_2}^m(\cos \theta_{b})
\;a_{n_3}(\zeta_3r_{1b}).
\end{split}
\end{align}
In the second expression we changed the variables from the Cartesian coordinates to the internal coordinate system
$(r_a,r_b,\phi)$ and integrated over the angle. Note, that the notation for the nonlinear parameters and for the
variable $n_3$ was suppressed since these quantities do not change during the recursive process. 
We have to
stress that all formulae presented here are valid only for $m>0$. There is no need to consider the negative values of
$m$ because of the axial symmetry of the integrands.

Generally speaking, to establish a recursive process which is able to increase the values of $l_1$, $l_2$ and $m$,
starting with provided values of $\langle \varphi_{n_1}^{00}|\varphi_{n_2}^{00}\rangle$ we need to use the 
well-known recursion relations for the Legendre polynomials $P_l^m$. A similar idea was applied by several authors to 
the calculation of various important matrix elements \cite{guseinov05,guseinov07,rico88,rico89b,rico93,rico91,rico89c}. 
Let us first derive a recursion relation connecting $\langle
\varphi_{n_1}^{mm}|\varphi_{n_2}^{mm}\rangle$ with different $m$ by recalling the following expression for the
Legendre polynomials with $l=m$:
\begin{align}
P_m^m(\cos \theta) = \frac{(2m)!}{2^m m!} \sin^m \theta,
\end{align}
so that
\begin{align}
P_{m+1}^{m+1}(\cos \theta) = P_{m}^{m}(\cos \theta) (2m+1)\sin \theta.
\end{align}
By combining two expressions like the above for $\cos \theta_a$ and $\cos \theta_b$ and using the obvious
relationship $r_a \sin \theta_a = r_b \sin \theta_b $ one finds 
\begin{align}
\begin{split}
&P_{m+1}^{m+1}(\cos \theta_a) P_{m+1}^{m+1}(\cos \theta_b)=\\
&P_{m}^{m}(\cos \theta_a) P_{m}^{m}(\cos \theta_b)
(2m+1)^2 \frac{r_a}{r_b} \sin^2 \theta_a,
\end{split}
\end{align}
and the expression for $\sin^2 \theta_a$ in terms of $r_a$, $r_b$ is elementary. Finally, this
leads to the recursion relation for the desired set of integrals
\begin{align}
\begin{split}
&\langle \varphi_{n_1}^{m+1,m+1}|\varphi_{n_2}^{m+1,m+1}\rangle = \frac{(2m+1)^2}{2R^2} \times \\
&\bigg[ R^2 \langle \varphi_{n_1+1}^{mm}|\varphi_{n_2-1}^{mm}\rangle
+R^2\langle \varphi_{n_1-1}^{mm}|\varphi_{n_2+1}^{mm}\rangle\\
&+\langle \varphi_{n_1+1}^{mm}|\varphi_{n_2+1}^{mm}\rangle-
\frac{1}{2}R^4\langle \varphi_{n_1-1}^{mm}|\varphi_{n_2-1}^{mm}\rangle\\
&-\frac{1}{2}\langle \varphi_{n_1+3}^{mm}|\varphi_{n_2-1}^{mm}\rangle-
\frac{1}{2}\langle \varphi_{n_1-1}^{mm}|\varphi_{n_2+3}^{mm}\rangle \bigg].
\end{split}
\end{align}

The second ingredient of the recursive process is a relation that allows to increase the values of $l_1$ and $l_2$
independently, starting with the just considered $\langle \varphi_{n_1}^{mm}|\varphi_{n_2}^{mm}\rangle$ integrals.
The following recursion relation for the Legendre polynomials is useful
\begin{align}
(l-m+1)P_{l+1}^m(x)+(l+m)P_{l-1}^m(x)=(2l+1)xP_l^m(x).
\end{align}
If one uses the above relation for $P_{l_1}^m(\cos \theta_a)$ in Eq. (\ref{pertinent1}) and subsequently expresses $\cos
\theta_a$ through $r_a$ and $r_b$ from the cosine theorem, the following recursion is obtained
\begin{align}
\label{rec1}
\begin{split}
&\frac{1}{2R}\bigg[
\langle \varphi_{n_1+1}^{l_1m}|\varphi_{n_2}^{l_2m}\rangle-
\langle \varphi_{n_1-1}^{l_1m}|\varphi_{n_2+2}^{l_2m}\rangle
+R^2\langle \varphi_{n_1-1}^{l_1m}|\varphi_{n_2}^{l_2m}\rangle \bigg]\\
&=(l_1-m+1) \langle \varphi_{n_1}^{l_1+1,m}|\varphi_{n_2}^{l_2m}\rangle
+(l_1+m) \langle \varphi_{n_1}^{l_1-1,m}|\varphi_{n_2}^{l_2m}\rangle,
\end{split}
\end{align}
which can be used to increase $l_1$ at cost of $n_1$ and $n_2$. A corresponding expression for increasing $l_2$ can be
obtained by repeating the derivation for $P_{l_2}^m(\cos \theta_b)$. Therefore, by using Eq. (\ref{rec1}) and its
counterpart for the centre $b$, we can build all $\langle \varphi_{n_1}^{l_1m}|\varphi_{n_2}^{l_2m}\rangle$ starting
with integrals with $l_1=l_2=m$ and higher $n_1$, $n_2$.

Having said this, the only thing that remains in question is the calculation of the pertinent integrals $\langle
n_100|n_200\rangle$. Let us return to Eq. (\ref{pertinent1})
\begin{align}
\begin{split}
\label{pertinent2}
\langle \varphi_{n_1}^{00}|\varphi_{n_2}^{00}\rangle &=
\int_0^\infty dr_a \int_{|r_a-R|}^{r_a+R} dr_b \;r_a^{n_1} r_b^{n_2}\\
&\times e^{-\zeta_1r_a-\zeta_2r_b}\;a_{n_3}(\zeta_3r_{1b}),
\end{split}
\end{align}
use the integral representation of $a_n$, Eq. (\ref{ap}), and reverse the order of integration. By doing so we obtain
an equivalent representation of the basic integrals
\begin{align}
\label{pertinent3}
\langle \varphi_{n_1}^{00}|\varphi_{n_2}^{00}\rangle =
\int_0^1 dt\; t^{n_3}\; \Gamma_{n_1n_2}(R;\zeta_1,\zeta_2+t\zeta_3),
\end{align}
where $\Gamma_{mn}$ are the usual overlap integrals between $ns$-type orbitals
\begin{align}
\label{tmn}
\Gamma_{mn}(R;\zeta_1,\zeta_2) = \int_0^\infty dr_a \int_{|r_a-R|}^{r_a+R} dr_b \;r_a^m r_b^n\;
e^{-\zeta_1r_a-\zeta_2r_b}.
\end{align}
In our approach, the outer integral in (\ref{pertinent3}) is carried out numerically. The arguments for this approach
are virtually the same as in the ellipsoidal coordinates method. Roughly speaking, numerical integration serves as a way
to avoid numerical instabilities which inevitably appear when the analytic approaches are used. However, now we require
a robust scheme for the calculation of $\Gamma_{mn}$, so that these integrals can be computed at each
point of the grid without a great overhead. In fact, the main advantage of the numerical integration in the ellipsoidal
coordinates method was that the integrand in Eq. (\ref{jl2}) could be evaluated extremely
efficiently and with a strictly controlled precision. On the other hand, the desired algorithm has to preserve a decent
accuracy up to large values of $m$ and $n$ (several tens, say). Determination of such an algorithm still presents a
challenge from the practical point of view.

\begin{table*}[ht]
\caption{Calculation of $\widetilde{S}_{13,12,m}^{15,12,m}({26};\zeta_1,0,\zeta_3)$ using the recursive method. The
values of $\zeta_1$, $\zeta_3$ increase along the columns or rows, respectively. The values presented
are in the form d-q which denotes (rounded) decimal logarithms of the relative error obtained in double and
quadruple arithmetic precision, respectively. Therefore, this values roughly represent the number of correct significant
digits obtained using the present algorithm. Values obtained in quadruple precision were demoted to double precision
before the comparison since these are the values used in the actual calculations. The worst result obtained within the
possible range of $m$ was chosen in all cases. Reference values were obtained from calculations in extended arithmetic
precision of 128 significant digits. }
\begin{ruledtabular}
\begin{tabular}{c|ccccccccccccc}
\label{table3}
$\zeta_1$/$\zeta_3$ & 0.1250 & 0.2500 & 0.5000 & 1.0000 & 2.0000 & 4.0000 & 8.0000 & 16.000 & 32.000 & 64.000 & 128.00 &
256.00\\
\hline\\[-2.1ex]
0.1250 & 0$-$0 & 0$-$0 & 0$-$0 & 0$-$0 & 0$-$0 & 0$-$5 & 0$-$11 & 1$-$16 & 4$-$16 & 5$-$16 & 1$-$16 & 0$-$11 \\[0.6ex]
0.2500 & 0$-$0 & 0$-$0 & 0$-$0 & 0$-$0 & 0$-$0 & 0$-$5 & 0$-$11 & 1$-$16 & 4$-$16 & 5$-$16 & 1$-$16 & 0$-$11 \\[0.6ex]
0.5000 & 0$-$0 & 0$-$0 & 0$-$0 & 0$-$0 & 0$-$0 & 0$-$5 & 0$-$11 & 1$-$16 & 4$-$16 & 5$-$16 & 1$-$16 & 0$-$11 \\[0.6ex]
1.0000 & 0$-$0 & 0$-$0 & 0$-$0 & 0$-$0 & 0$-$0 & 0$-$5 & 0$-$11 & 1$-$16 & 4$-$16 & 5$-$16 & 1$-$16 & 0$-$11 \\[0.6ex]
2.0000 & 0$-$0 & 0$-$0 & 0$-$0 & 0$-$0 & 0$-$0 & 0$-$6 & 0$-$10 & 0$-$14 & 4$-$16 & 4$-$16 & 0$-$14 & 0$-$11 \\[0.6ex]
4.0000 & 0$-$6 & 0$-$6 & 0$-$6 & 0$-$6 & 0$-$6 & 0$-$5 & 0$-$11 & 1$-$16 & 3$-$16 & 4$-$16 & 1$-$16 & 0$-$12 \\[0.6ex]
8.0000 & 0$-$12 & 0$-$12 & 0$-$12 & 0$-$12 & 0$-$10 & 0$-$12 & 1$-$16 & 3$-$16 & 3$-$16 & 2$-$16 & 1$-$16 & 0$-$12
\\[0.6ex]
16.000 & 1$-$16 & 1$-$16 & 1$-$16 & 1$-$16 & 0$-$14 & 1$-$16 & 1$-$16 & 2$-$16 & 2$-$16 & 1$-$16 & 0$-$15 & 0$-$11
\\[0.6ex]
32.000 & 4$-$16 & 4$-$16 & 4$-$16 & 4$-$16 & 4$-$16 & 3$-$16 & 3$-$16 & 5$-$16 & 6$-$16 & 2$-$16 & 1$-$16 & 0$-$14
\\[0.6ex]
64.000 & 5$-$16 & 5$-$16 & 5$-$16 & 5$-$16 & 4$-$16 & 3$-$16 & 4$-$16 & 7$-$16 & 9$-$16 & 6$-$16 & 1$-$16 & 0$-$15
\\[0.6ex]
128.00 & 1$-$16 & 1$-$16 & 1$-$16 & 1$-$16 & 0$-$14 & 1$-$16 & 1$-$16 & 1$-$16 & 4$-$16 & 3$-$16 & 0$-$16 & 0$-$13
\\[0.6ex]
256.00 & 0$-$12 & 0$-$12 & 0$-$12 & 0$-$12 & 0$-$10 & 0$-$12 & 0$-$12 & 0$-$12 & 0$-$13 & 0$-$13 & 0$-$14 & 0$-$12
\\[0.6ex]
\end{tabular}
\end{ruledtabular}
\end{table*}

The basic integrals $\Gamma_{mn}$ are well-known in the literature. Many authors considered their computation by using
several different algorithms which varied in accuracy and speed. Let us note, however, that in the calculation of the
integrals (\ref{tmn}) the main issue is the numerical stability. The actual expressions for these integrals are not
difficult to derive and include only simple elementary functions. Unfortunately, these expressions consist of finite
series with terms of alternating signs. When $m$, $n$ are increased these terms grow exponentially while the sum
remains by orders of magnitude smaller. As a result, a gross digital erosion is inevitable.
In a large fraction of works which considered calculation of the integrals (\ref{tmn}), or used them as a part of
different algorithms, the issue of numerical stability was completely disregarded or treated very lightly. The common
justification for this fact is that authors were mainly interested in low quantum numbers or devised their
algorithms to verify the correctness of the approach more than to perform general calculations.

Let us begin by noting that all integrals (\ref{tmn}) can be generated by a consecutive differentiation of $\Gamma_{00}$
with respect to the nonlinear parameters $\zeta_1$, $\zeta_2$ \emph{i.e.}
\begin{align}
\label{Gmn}
\Gamma_{mn}(R;\zeta_1,\zeta_2) = 
\left(-\frac{\partial}{\partial \zeta_1}\right)^m
\left(-\frac{\partial}{\partial \zeta_2}\right)^n \Gamma_{00}(\zeta_1,\zeta_2),
\end{align}
which is, in substance, a trivial case of the so-called shift method of Fern\'{a}ndez Rico \emph{et al.} 
\cite{rico00a,rico00b,rico01}. The simplest integrals $\Gamma_{00}$ are elementary
\begin{align}
\label{G00}
\Gamma_{00}(R;\zeta_1,\zeta_2) = \frac{2}{\zeta_1+\zeta_2} \frac{e^{-\zeta_2R}-e^{-\zeta_1R}}{\zeta_1-\zeta_2}.
\end{align}
It is now convenient to define $g_{00}$ by
\begin{align}
\label{g00}
 g_{00}(R;\zeta_1,\zeta_2) = 2\,\frac{e^{-\zeta_2R}-e^{-\zeta_1R}}{\zeta_1-\zeta_2},
\end{align}
so that $\Gamma_{00}=g_{00}/(\zeta_1+\zeta_2)$, and the definition of $g_{mn}$ is analogous
\begin{align}
\label{gmn}
g_{mn}(R;\zeta_1,\zeta_2) = 
\left(-\frac{\partial}{\partial \zeta_1}\right)^m
\left(-\frac{\partial}{\partial \zeta_2}\right)^n g_{00}(\zeta_1,\zeta_2).
\end{align}
Let us now multiply both sides of Eq. (\ref{G00}) by $\zeta_1+\zeta_2$, rewrite the result in terms of $g_{00}$ and
differentiate both sides $m$ with respect to $-\zeta_1$ and $n$ times with respect to $-\zeta_2$. After some
rearrangements, the final result can be written as
\begin{align}
\label{Grec}
\Gamma_{mn} = \frac{1}{\zeta_1+\zeta_2}\bigg[g_{mn}+m\Gamma_{m-1,n}+n\Gamma_{m,n-1}\bigg],
\end{align}
where the notation for the nonlinear parameters is suppressed for brevity. The above expression is an inhomogeneous
linear recursion relation for $\Gamma_{mn}$. Note, that all integrals $\Gamma_{mn}$ are positive and so are the values 
of $g_{mn}$. Therefore, the above recursion relation is completely stable. This approach is reminiscent of the 
treatment of the one-centre integrals by Sack \emph{et al.} \cite{sack67}.

\begin{table*}[ht]
\caption{Calculation of $\widetilde{S}_{7,6,m}^{21,18,m}({26};\zeta_1,\zeta_1,\zeta_3)$ using the recursive method. The
values of $\zeta_1$, $\zeta_3$ increase along the columns or rows, respectively. The values presented
are in the form d$-$q which denotes (rounded) decimal logarithms of the relative error obtained in double and
quadruple arithmetic precision, respectively. Therefore, this values roughly represent the number of correct significant
digits obtained using the present algorithm. Values obtained in quadruple precision were demoted to double precision
before the comparison since these are the values used in the actual calculations. The worst result obtained within the
possible range of $m$ was chosen in all cases. Reference values were obtained from calculations in extended arithmetic
precision of 128 significant digits. }
\begin{ruledtabular}
\begin{tabular}{c|ccccccccccccc}
\label{table4}
$\zeta_1$/$\zeta_3$ & 0.1250 & 0.2500 & 0.5000 & 1.0000 & 2.0000 & 4.0000 & 8.0000 & 16.000 & 32.000 & 64.000 & 128.00 &
256.00\\
\hline\\[-2.1ex]
0.1250 & 0$-$0 & 0$-$0 & 0$-$0 & 0$-$0 & 0$-$0 & 0$-$0 & 0$-$0 & 0$-$0 & 0$-$0 & 0$-$0 & 0$-$0 & 0$-$0 \\[0.6ex]
0.2500 & 0$-$0 & 0$-$0 & 0$-$0 & 0$-$0 & 0$-$0 & 0$-$0 & 0$-$0 & 0$-$0 & 0$-$0 & 0$-$0 & 0$-$0 & 0$-$0 \\[0.6ex]
0.5000 & 0$-$0 & 0$-$0 & 0$-$0 & 0$-$0 & 0$-$0 & 0$-$0 & 0$-$0 & 0$-$0 & 0$-$0 & 0$-$0 & 0$-$0 & 0$-$0 \\[0.6ex]
1.0000 & 0$-$0 & 0$-$0 & 0$-$0 & 0$-$0 & 0$-$0 & 0$-$0 & 0$-$0 & 0$-$0 & 0$-$0 & 0$-$0 & 0$-$0 & 0$-$0 \\[0.6ex]
2.0000 & 0$-$0 & 0$-$0 & 0$-$0 & 0$-$0 & 0$-$0 & 0$-$0 & 0$-$0 & 0$-$0 & 0$-$0 & 0$-$0 & 0$-$0 & 0$-$0 \\[0.6ex]
4.0000 & 0$-$0 & 0$-$0 & 0$-$0 & 0$-$0 & 0$-$0 & 0$-$0 & 0$-$0 & 0$-$0 & 0$-$0 & 0$-$0 & 0$-$0 & 0$-$0 \\[0.6ex]
8.0000 & 0$-$5 & 0$-$5 & 0$-$5 & 0$-$5 & 0$-$5 & 0$-$5 & 0$-$5 & 0$-$6 & 0$-$6 & 0$-$6 & 0$-$5 & 0$-$4
\\[0.6ex]
16.000 & 0$-$12 & 0$-$12 & 0$-$12 & 0$-$12 & 0$-$12 & 0$-$12 & 0$-$12 & 0$-$13 & 0$-$13 & 0$-$13 & 0$-$12 & 0$-$11
\\[0.6ex]
32.000 & 6$-$16 & 6$-$16 & 6$-$16 & 6$-$16 & 6$-$16 & 6$-$16 & 6$-$16 & 6$-$16 & 7$-$16 & 7$-$16 & 4$-$16 & 0$-$14
\\[0.6ex]
64.000 & 8$-$16 & 8$-$16 & 8$-$16 & 8$-$16 & 8$-$16 & 8$-$16 & 8$-$16 & 8$-$16 & 9$-$16 & 10$-$16 & 6$-$16 & 0$-$15
\\[0.6ex]
128.00 & 4$-$16 & 4$-$16 & 4$-$16 & 4$-$16 & 4$-$16 & 4$-$16 & 4$-$16 & 4$-$16 & 6$-$16 & 7$-$16 & 2$-$16 & 0$-$14
\\[0.6ex]
\end{tabular}
\end{ruledtabular}
\end{table*}

The problem is now reduced to an efficient calculation of $g_{mn}$. Explicit differentiation is not an option
because of similar cancellations as for the initial $\Gamma_{mn}$ integrals. However, let us observe that $g_{00}$ can
also be rewritten as
\begin{align}
g_{00}(\zeta_1,\zeta_2) = \frac{R}{2} e^{-\zeta_1R} M\big[1,2,(\zeta_1-\zeta_2)R\big],
\end{align}
where $M(a,b,z)$ is the confluent hypergeometric function \cite{stegun72} (denoted as $_1F_1$ by some authors). By 
using 
two
differentiation formulae for $M(a,b,z)$
\begin{align}
\frac{\partial^n}{\partial z^n} M(a,b,z) &= \frac{(a)_n}{(b)_n} M(a+n,b+n,z),\\
\frac{\partial^n}{\partial z^n} \bigg[ e^{-z} M(a,b,z) \bigg] &= (-1)^n \frac{(b-a)_n}{(b)_n} e^{-z} M(a,b+n,z),
\end{align}
one easily arrives at the new formula for $g_{mn}$
\begin{align}
\begin{split}
g_{mn}(\zeta_1,\zeta_2) &= \frac{1}{2}\;e^{-\zeta_1R}\;R^{m+n+1}\times\\
&\times M\big[1+n,2+m+n,(\zeta_1-\zeta_2)R\big].
\end{split}
\end{align}
At this point the problem can be considered to be solved because methods of calculation of $M(a,b,z)$ for arbitrary real
(or even complex) values of the parameters $a$, $b$, and $z$ exist. Let us note that here we deal with
an exceptionally special case of $M(a,b,z)$ with both $a$ and $b$ being strictly positive integers, and additionally
$b>a$ always holds. Moreover, we can use the symmetry of the initial integrals, $\langle
\varphi_{n_1}^{00}|\varphi_{n_2}^{00}\rangle = \langle \varphi_{n_2}^{00}|\varphi_{n_1}^{00} \rangle$, in order to
impose the restriction $\zeta_1\geq\zeta_2$, which gives $z\geq 0$. All these conditions signal that we should design a
dedicated procedure for the calculation of $M(a,b,z)$ in this special case and avoid using general algorithms which are
drastically more complicated and involve a large computational overhead. In Appendix we present a recursive 
method
which is able to calculate $M(a,b,z)$ in our special case with a decent speed, at the same time preserving full accuracy
allowed by the arithmetic.

In Tables 3 and 4 we present results of the benchmark calculations for the same representative integrals, 
$\widetilde{S}_{1312m}^{1512m}({26};\zeta_1,0,\zeta_3)$ and 
$\widetilde{S}_{7,6,m}^{21,18,m}({26};\zeta_1,\zeta_1,\zeta_3)$, as in the previous subsection. We use the same 
numerical quadrature as before and typically a machine precision is obtained in Eq. (\ref{pertinent3}). One sees that 
the recursive algorithm fails completely, even in the quadruple arithmetic precision, when nonlinear parameters are 
both small. On the other hand, as they get large the accuracy gradually improves which is exactly the opposite 
behaviour to the one found in the ellipsoidal method. Therefore, two methods presented in this paper can be 
considered fully complementary and together are able to cover a sufficiently large range of the nonlinear parameters. 
Outside this range, hybrid integrals are usually very small and are typically neglected in advance by the Schwarz 
screening technique or a similar scheme. Coulomb integrals with bigger values of the nonlinear parameters may still be 
non-negligible. However, they can be computed with different standard techniques such as the multipole expansion. 
It is mandatory for a general program to include such a method as an option.

\section{Conclusions}

Concluding, we derived new expressions for the Coulomb and hybrid integrals over the Slater-type orbitals, with no
restrictions on the values of the quantum numbers, starting by a direct integration over coordinates of the second 
electron.
In this way the desired integrals reduce to combinations of ordinary overlap integrals and a set of the so-called
overlap-like integrals. These basic integrals are evaluated by using two distinct methods - direct integration in
the ellipsoidal coordinate system or with a recursive scheme for increasing angular momenta in the integrand. One of
the biggest problems in actual computations is numerical stability of the resulting algorithms. Many formulations
available in the literature contain numerically badly conditioned expressions which introduce a significant loss of
digits when evaluated in a finite arithmetic precision. We show how these instabilities can be avoided if a 
simple, one-dimensional numerical integration is used instead. We discuss that this numerical approach introduces an
acceptable computational overhead due to well-behaved and simple form of the integrands. We also show that the
remaining numerical instabilities can be easily controlled. Extensive numerical tests are presented, verifying the
usefulness and applicability of the method.

\begin{acknowledgments}
This work was supported by the Polish Ministry of Science and Higher Education, grant NN204 182840. ML acknowledges the
Polish Ministry of Science and Higher Education for the support through the project \textit{``Diamentowy Grant''},
number DI2011 012041. RM was supported by the Foundation for Polish Science through the \textit{``Mistrz''} program. We
would like to thank Bogumi\l~Jeziorski for fruitful discussions, reading and commenting on the manuscript.
\end{acknowledgments}

\appendix*

\section{Calculation of $M(a,b,z)$ for $a,b \in \mathbb{Z}_+$, $b>a$, $z\geq 0$}
\label{app:appa}

Let us start by recalling some of the useful formulae obeyed by $M(a,b,z)$. The first one is the Gautschi representation
of the continued fraction (GCF) \cite{gautschi77} which states that
\begin{align}
\label{gcf}
\begin{split}
&\frac{M(a+1,b,z)}{M(a,b,z)} = 1 + \frac{z}{a}\sum_{k=0}^\infty p_k,\\
&p_0 = 1,\;\;\; p_k = \prod_{i=1}^k r_i,\\
&r_0 = 0,\;\;\; r_k = -\frac{a_k(1+r_{k-1})}{1+a_k(1+r_{k-1})},\\
&a_k = \frac{(a+k)z}{(b-z+k-1)(b-z+k)}.
\end{split}
\end{align}
The second useful expression is the recursion relation which allows to increase the value of $a$ at constant $b$:
\begin{align}
\label{recm}
\begin{split}
&(b-a)M(a-1,b,z)+(2a-b+z)M(a,b,z)\\
&-aM(a+1,b,z)=0.
\end{split}
\end{align}
The region $a,b \in \mathbb{Z}_+$, $b>a$, $z\geq 0$ needs to be divided into three subregions and different algorithms
have to be used in each of them. They are as follows:
\begin{itemize}
\item $b\geq2a+z$,\\
one first uses GCF, Eq. (\ref{gcf}), in order to obtain the ratio $M(a+1,b,z)/M(a,b,z)$ for the maximal desired $b$ and
$a=\lceil(b-z)/2\rceil$ ($\lceil * \rceil$ is the ceiling function). The recursion (\ref{recm}) can be rewritten as
\begin{align}
\label{miller}
r_{a-1} = \frac{b-a}{ar_a+b-2a-z},
\end{align}
where $r_a = M(a+1,b,z)/M(a,b,z)$. This recursion is then carried out downward, starting with the value of the ratio
obtained from GCF, until $r_0$ is reached. Since $M(0,b,z)=1$, it turns out that $r_0 = M(1,b,z)$ and other values can
be obtained by using the definition of $r_a$ \emph{e.g.} $M(2,b,z) = r_1 M(1,b,z)$.
\item $b<2a+z$, $b\geq z$,\\
again, the relation (\ref{recm}) is transformed into a Miller-like two-step recursion
\begin{align}
r_a=\frac{b-a}{a}\frac{1}{r_{a-1}}+2+\frac{z-b}{a},
\end{align}
with $r_a$ being defined in the same way as previously.
Starting with an arbitrary value of $r_0$, this recursion is carried out upward up to the line $a=b$ (corresponding to
$r_{b-1}$). Using the exact relationship $M(b,b,z)=e^z$ one finds that actual values of $M(a,b,z)$ can be reconstructed
as $M(b-1,b,z)=M(b,b,z)/r_{b-1}=e^z/r_{b-1}$, $M(b-2,b,z)=M(b-1,b,z)/r_{b-2}$ \emph{etc.} until the value of $M(1,b,z)$
is reached.
\item $b<2a+z$, $b<z$,\\
this is the so-called anomalous convergence region of GCF \emph{i.e.} the expression (\ref{gcf}) converges to the wrong
result \cite{gautschi77} and therefore cannot be used. However, in this region the initial upward recursion (\ref{recm})
is totally stable since all terms in (\ref{recm}) are positive. The starting (exact) values are
\begin{align}
M(0,b,z) &= 1,\\
M(1,b,z) &= (b-1)e^z a_{b-2}(z),
\end{align}
where $a_n$ are given by Eq. (\ref{ap}). The second relationship breaks down when $b=1$ but in this case we obtain
independently $M(1,1,z)=e^z$, as noted beforehand.
\end{itemize}
Let us also add in passing that the power series expansion of $M(a,b,z)$ around $z=0$ can additionally be used for small
$z$ 
\begin{align}
 M(a,b,z) = \sum_{s=0}^\infty \frac{(a)_s}{(b)_s s!}z^s,
\end{align}
since it typically converges very fast in the vicinity of the origin, $z\approx 0$. Similar conclusion holds for the
asymptotic
expansion of $M(a,b,z)$ as $z$ is large. Remarkably, when the values of $M(a,b,z)$ are calculated as described in this
Appendix, no loss of digits is observed, and thus $\langle \varphi_{n_1}^{00}|\varphi_{n_2}^{00}\rangle$ can be obtained
with
full precision up to very large values of $n_1$ and $n_2$.

\end{document}


\title{{\Large Supplemental Material.} \\[0.5em] \noindent 
Calculation of two-centre two-electron integrals over Slater-type orbitals revisited. \\I.
Coulomb and hybrid integrals }

\author{\sc Micha\l\ Lesiuk\footnote[1]{e-mail: lesiuk@tiger.chem.uw.edu.pl} 
and Robert Moszynski}
\affiliation{\sl Faculty of Chemistry, University of Warsaw\\
Pasteura 1, 02-093 Warsaw, Poland}
\date{\today}

\begin{abstract}
The present document serves as a supplemental material for the publication \emph{Calculation of two-centre
two-electron integrals over Slater-type orbitals revisited. I. Coulomb and hybrid integrals}. It contains data or
derivations which are of some importance but are not necessary for an overall understanding of the manuscript. However,
the material presented here will be useful for a reader who wishes to repeat all of our derivations in details.
Researchers who wish to repeat some of the calculations independently may also benefit from these additional data.
\end{abstract}

\maketitle

\section{Supplemental material for Paper I}

Let us consider an atomic integral
\begin{align}
\begin{split}
I = \int d\textbf{r}_1 \int d\textbf{r}_2\,
\chi_{n_1l_1}(\textbf{r}_1;\zeta_1)
\chi_{n_2l_2}(\textbf{r}_1;\zeta_2)
\frac{1}{r_{12}}
\chi_{n_3l_3}(\textbf{r}_2;\zeta_3)
\chi_{n_4l_4}(\textbf{r}_2;\zeta_4),
\end{split}
\end{align}
where all orbitals are located at the same point of space. By using the Laplace expansion for the term
$\frac{1}{r_{12}}$ and integrating over the angular coordinates of both electrons one gets
\begin{align}
\small
\begin{split}
I &= (-1)^{m_1+m_3} \sqrt{\prod_{i=1}^4 (2l_i+1)}
\sum_L \wigner{l_1}{L}{l_2}{0}{0}{0} \wigner{l_3}{L}{l_4}{0}{0}{0}\\
& \times \Big[ I_L^>(n_{12}+2,n_{34}+2;\zeta_{12},\zeta_{34}) + I_L^<(n_{12}+2,n_{34}+2;\zeta_{12},\zeta_{34}) \Big] \\
& \times \sum_{M=-L}^L (-1)^M \wigner{l_1}{L}{l_2}{-m_1}{-M}{m_2} \wigner{l_3}{L}{l_4}{-m_3}{M}{m_4},
\end{split}
\end{align}
where $n_{12}=n_1+n_2$, $\zeta_{12}=\zeta_1+\zeta_2$ and the same notation is used for the second electron. The
summation over $L$ in the above expression is finite \emph{i.e.} only terms with $\max(|l_1-l_2|,|l_3-l_4|)\leq L \leq
\min(l_1+l_2,l_3+l_4)$ survive. Let us first consider the integral $I_L^>$ which takes the form
\begin{align}
\begin{split}
I_L^>(n_1,n_2;\alpha_1,\alpha_2) &= \int_0^\infty dr_1\, r_1^{n_1+L} e^{-\alpha_1 r_1}
\int_{r_1}^\infty dr_2\,r_2^{n_2-L-1} e^{-\alpha_2 r_2},
\end{split}
\end{align}
where the notation for the parameters was simplified for better readability. This integral can be integrated by using
elementary methods to give
\begin{align}
\begin{split}
I_L^>(n_1,n_2;\alpha_1,\alpha_2) = \frac{\alpha_2^{L-n_2}(n_2-L-1)!}{(\alpha_1+\alpha_2)^{n_1+L+1}}
\times \sum_{j=0}^{n_2-L-1}\frac{(n_1+L+j)!}{j!} \left( \frac{\alpha_2}{\alpha_1+\alpha_2} \right)^j.
\end{split}
\end{align}
Note all terms present in the above sum are positive and thus no cancellation of huge numbers to a relatively small
result can occur. The above expression can be put even in a more compact form by using the Pochhammer symbols.

Let us now pass to the second class of integrals ($I_L^<$) which are defined as
\begin{align}
\label{ismall}
I_L^<(n_1,n_2;\alpha_1,\alpha_2) &= \int_0^\infty dr_1\, r_1^{n_1-L-1} e^{-\alpha_1 r_1}
\int_{0}^{r_1} dr_2\,r_2^{n_2+L} e^{-\alpha_2 r_2}.
\end{align}
In some works, the first step to bring the above integrals into a closed form is to manipulate the integration range in
the inner integral by inserting an obvious identity $\int_{0}^{r_1} = \int_{0}^{\infty} - \int_{r_1}^{\infty}$. The
advantage of this idea is that all resulting integrals are solved immediately in the same way as $I_L^>$. However, for
some combinations of $\alpha_2$ and $r_1$ the two resulting integrals are nearly equal in magnitude. Therefore,
subtraction will cause a significant loss of accuracy. This effect is particularly considerable when the values of the
quantum numbers are large.

In the alternative approach, the integration variable in the inner
integrand of Eq. (\ref{ismall}) is changed to $t=r_2/r_1$. If the order of integration is reversed, the integration
over $r_1$ can be carried out easily and one is left with the following one-dimensional integration
\begin{align}
I_L^<(n_1,n_2;\alpha_1,\alpha_2) = (n_1+n_2)! \int_0^1 dt\, \frac{t^{n_2+L}}{(\alpha_1+\alpha_2t)^{n_1+n_2+1}}.
\end{align}
The above integrand is a rational function and the integration can be carried out by using general methods. Taking into
account that the relationship $L\leq n_2$ always holds, the final result can be cast into the form
\begin{align}
\begin{split}
I_L^<(n_1,n_2;\alpha_1,\alpha_2) &= \frac{1}{\alpha_1} \frac{(n_2+L)!(n_1-L-1)!}{(\alpha_1+\alpha_2)^{n_1+n_2}}
\sum_{k=0}^{n_1-L-1} { n_1+n_2 \choose n_1-L-1-k } \left( \frac{\alpha_2}{\alpha_1} \right)^k.
\end{split}
\end{align}
The above summation includes only terms with positive signs and therefore the whole procedure is completely stable.
Moreover, no special functions or infinite summations are required as contrasted to some of the formulations available
in the literature.